\documentclass[10pt,fleqn]{article}

\usepackage{amssymb}

\newcommand{\name}[1]{\begin{flushleft}
                       \LARGE \bf #1
                       \end{flushleft}\vspace{-3mm}}

\newcommand{\Author}[1]{\begin{flushleft}
                       \it #1 \end{flushleft}}

\newcommand{\Adress}[1]{\begin{flushleft}
                       \it #1 \end{flushleft}}

\newcommand{\AbsEng}[1]{
    \begin{flushright}
    \begin{minipage}{120mm}
     \small   #1
    \end{minipage}
    \end{flushright}
}

\newcommand{\be}{\begin{equation}}
\newcommand{\ee}{\end{equation}}
\newcommand{\ba}{\hspace*{-5pt}\begin{array}}
\newcommand{\ea}{\end{array}}
\newcommand{\p}{\partial}
\newcommand{\ds}{\displaystyle}

\newcommand{\pbf}[1]{\mbox{\mathversion{bold}$#1$}}

\begin{document}

\name{On the $\pbf{P}$- and $\pbf{T}$-non-invariant two-component equation for
the neutrino}

\medskip

\noindent{published in {\it Nuclear Physics B}, 1970,  {\bf 21}, P. 321--330.}

\Author{Wilhelm I. FUSHCHYCH\par}

\Adress{Institute of Mathematics of the National Academy of
Sciences of Ukraine, \\ 3 Tereshchenkivska  Street, 01601 Kyiv-4,
UKRAINE}

\noindent {\tt URL:
http://www.imath.kiev.ua/\~{}appmath/wif.html\\ E-mail:
symmetry@imath.kiev.ua}

\AbsEng{The relativistic two-component equation describing the free motion of particles with zero
mass and spin $\frac 12$, which is $P$- and $T$-non-invariant but $C$-invariant, is found.
The representation of the Poincar\'e group for zero mass and discrete spin is constructed.
The position operator for such a particle is defined.}

\medskip

\renewcommand{\theequation}{1.\arabic{equation}}
\setcounter{equation}{0}

\centerline{\bf 1. Introduction}

As is known, the Dirac equation for a particle with zero mass:
\be
i\frac{\p \Psi (t,\pbf{x})}{\p t} =\gamma_0 \gamma_k p_k \Psi(t,\pbf{x}), \qquad k=1,2,3,
\ee
is invariant with respect to the space-time reflections. If one chooses for the Dirac matrices
the Weyl representation eq.~(1.1) decomposes into a system of two equations
\be
i\frac{\p \Psi_\pm(t,\pbf{x})}{\p t} =\pm \sigma_k p_k \Psi_\pm (t,\pbf{x}),
\ee
where $\sigma_k$ are the Pauli matrices and $\Psi_\pm$ is a two-component spinor. The Weyl
equation~(1.2) for $\Psi_+$ (or $\Psi_-$) is not invariant under space reflection
$P$ and charge conjugation $C$ but is invariant under the $CP$- and $T$-operations.

Due to the fact that the space parity in the weak interactions is not conserved it is usually
assumed that neutrino is described, not by the four-component eq.~(1.1), but by a two-component
one~(1.2). Therefore, in papers~[1] an hypothesis was put forward that the weak interactions
are invariant with respect to the $CP$ operation and consequently to the $T$ operation, if the
$CPT$ theorem is valid.

In this paper the two-component equation for a particle with zero mass and spin~$\frac 12$,
which is non-invariant under the time reflection of $T$ and the $CP$ operation, is found.

\medskip
\renewcommand{\theequation}{2.\arabic{equation}}
\setcounter{equation}{0}

\centerline{\bf 2. Equation for a neutrino with ``variable mass''}

On the solutions of eq. (1.1) the generators of the Poincar\'e group $P(1,3)$ have the form
\be
\ba{l}
\ds P_0^\Psi ={\mathcal H}^\Psi =\gamma_0 \gamma_k p_k, \qquad P_k^\Psi=p_k,
\vspace{1mm}\\
\ds J_{kl}^\Psi =x_k p_l -x_l p_k +S_{kl}, \qquad
J_{0k}^\Psi =x_0 p_k -\frac 12 \left[ x_k, {\mathcal H}^\Psi\right]_+,
\ea
\ee
\renewcommand{\theequation}{2.\arabic{equation}${}'$}
\setcounter{equation}{0}
\be
\ba{l}
\ds S_{\mu\nu} =\frac 14 i (\gamma_\mu \gamma_\nu -\gamma_\nu \gamma_\mu),
\qquad S_{\mu 4}=\frac 14 i (\gamma_\mu \gamma_4 -\gamma_4 \gamma_\mu),
\vspace{2mm}\\
\ds S_{\mu 5} =\frac 12 i\gamma_\mu, \qquad S_{45}=\frac 12 i\gamma_4,
\qquad \mu=0,1,2,3,
\ea
\ee
where  $\gamma_\mu$ and $\gamma_4$ are the Dirac matrices.

If one performs a unitary transformation [2] over eq. (1.1)
\renewcommand{\theequation}{2.\arabic{equation}}
\setcounter{equation}{1}
\be
U_1 =\exp \left\{ \frac 12 i \pi S_{53} e_3\right\}, \qquad e_3 =\frac{p_3}{|p_3|},
\qquad p_3\not= 0,
\ee
or
\renewcommand{\theequation}{2.\arabic{equation}${}'$}
\setcounter{equation}{1}
\be
U_1 =\frac{1}{\sqrt{2}} (1+\gamma_3 e_3),
\ee
eq.~(1.1) has the form
\renewcommand{\theequation}{2.\arabic{equation}}
\setcounter{equation}{2}
\be
i\frac{\p \chi (t,\pbf{x})}{\p t} =(\gamma_0 \gamma_a p_a +\gamma_0 |p_3|) \chi(t,\pbf{x}),
\qquad a=1,2,
\ee
\be
\chi=U_1 \Psi, \qquad \chi \equiv \left( \begin{array}{c} \chi_+ \\ \chi_-\end{array} \right),
\ee
where $\chi_\pm$ is a two-component spinor.

The Poincar\'e group generators  $P(1,3)$ on $\{\chi\}$ being the solution of eq.~(2.3) have
the form
\be
\ba{l}
\ds P_0^\chi ={\mathcal H}^\chi =\gamma_0 \gamma_a p_a +\gamma_0 |p_3|,
\qquad P_k^\chi =p_k,
\vspace{1mm}\\
\ds J_{ab}^\chi =x_a p_b -x_b p_a +S_{ab},
\qquad J_{a3}^\chi =x_a p_3 -x_3 p_a -e_3 S_{a3} \gamma_3,
\vspace{1mm}\\
\ds  J_{0k}^\chi =x_0 p_k -\frac 12 \left[x_k, {\mathcal H}^\chi\right]_+.
\ea
\ee
Choosing for the Dirac matrices somewhat unusual representation
\be
\ba{l}
\gamma_0 =\left( \begin{array}{cc}
\sigma_3 & 0\\
0 & -\sigma_3 \end{array}\right), \qquad
\gamma_a =\left( \begin{array}{cc}
i\sigma_a & 0\\
0 & -i\sigma_a \end{array}\right),
\vspace{3mm}\\
\gamma_3 =\left( \begin{array}{cc}
0 & i\\
i & 0 \end{array}\right), \qquad
\gamma_4 =\left( \begin{array}{cc}
0 & i\\
-i & 0 \end{array}\right),
\ea
\ee
eq. (2.3) decomposes into a system of two equations
\be
\ba{l}
\ds i\frac{\p \chi_\pm (t,\pbf{x})}{\p t} =\left\{ i\sigma_3 \sigma_a p_a \pm \sigma_3 |p_3|\right\}
\chi_\pm (t,\pbf{x}),
\vspace{2mm}\\
\ds \chi_\pm =Q_\pm \chi, \qquad
Q_\pm =\frac 12 \pm i S_{43} =\frac 12 (1\pm \gamma_3\gamma_4).
\ea
\ee

Eq. (2.7) for the functions $\chi_+(t,\pbf{x})$ (or $\chi_-(t,\pbf{x})$) has quite the other properties
relative to the discrete transformations than the Weyl equation~(1.2).

We note the following:

(i) It is possible to arrive at eq. (2.3) (or (2.7)) in another way. If we ``extract the square root''
from the operator equation
\[
\left( p_0^2-p_a^2\right) \chi =p_3^2 \chi,
\]
we obtain eqs. (2.3) (or eqs. (2.7)).

(ii) The fact that the Dirac equations for zero and non-zero mass are invariant under the
$P$-, $T$- and $C$-transformations is the consequence of the fact that they, besides
being invariant with respect to the group $P(1,3)$, are invariant under the group
$SU(2)\otimes SU(2) \sim O(4)$ (this question will be considered in detail in a following paper).

Eq. (2.3) coincides in form with a usual Dirac equation for zero mass if $|p_3|$ is considered
as the mass of a particle. Therefore it is possible to say that eq.~(2.3) describes a
``flat neutrino'' with variable mass $|p_3|$. Really the operator $|p_3|$ is the Casimir
operator of the group $P(1,2)$ but not of the group $P(1,3)$.

Before passing to an investigation of the $P$-, $T$- and $C$-properties of eqs.~(2.7) we shall
construct the operator of the position in the space.

For eq. (2.3) the operator of the Foldy--Wouthuysen type has the form
\be
U_2 =\exp \left\{ \frac{S_{5a} p_a}{\sqrt{p_a^2}} \, \mbox{arctg}\, \frac{\sqrt{p_a^2}}{|p_3|}\right\}.
\ee
If the matrix $S_{5a}$ have the form of ($2.1'$) then
\be
U_2 =\frac{E+|p_3| +\gamma_a p_a}{\{ 2E(E+|p_3|)\}^{1/2}},
\qquad E=\sqrt{p_1^2 +p_2^2 +p_3^2}.
\ee
Eq.~(2.3) after the transformation (2.9) transfers into
\be
i\frac{\p \Phi (t, \pbf{x})}{\p t} ={\mathcal H}^\Phi (t,\pbf{x})= \gamma_0 E\Phi(t,\pbf{x}),
\qquad \Phi(t,\pbf{x})=U_2 \chi(t,\pbf{x}).
\ee

The generators of the group $P(1,3)$ on $\{\Phi\}$ have the form
\be
\ba{l}
\ds P_0^\Phi ={\mathcal H}^\Phi =\gamma_0 E, \qquad P_k^\Phi =p_k,
\vspace{2mm}\\
\ds J_{ab}^\Phi =x_a p_b -x_b p_a +S_{ab}, \qquad
J_{a3}^\Phi =x_a p_3 -x_3 p_a -e_3 \frac{S_{ab} p_b}{E+|p_3|},
\vspace{2mm}\\
\ds J_{0a}^\Phi =x_0 p_a -\frac 12 \left[ x_a, {\mathcal H}^\Phi\right]_+ -
\gamma_0 \frac{S_{ab} p_b}{E+|p_3|}, \qquad
J_{03}^\Phi =x_0 p_3 -\frac 12 \left[ x_3 , {\mathcal H}^\Phi\right]_+.
\ea\!
\ee

It must be noted that the operators (2.11), as it can be immediately verified, satisfy the
algebra $P(1,3)$ commutation relations not depending on the matrices $S_{ab}$
explicit form, i.e. the operators~(2.11), if $\gamma_0$ is substituted for 1 (or $-1$)
and realize irreducibly the algebra $P(1,3)$ representation which is characterized by zero
mass and discrete spin. The representation~(2.11) differs from the corresponding Shirokov~[3],
Lomont--Moses~[4] ones but is certainly equivalent to them.

The position operator on a set $\{\chi\}$ looks as
\be
\ba{l}
\ds X_a^\chi =U_2^{-1} x_a U_2 =x_a -\frac{S_{5a}}{E} +\frac{S_{5c} p_c p_a}{E^2 (E+|p_3|)}
+\frac{S_{ac} p_c}{E(E+|p_3|)},
\vspace{3mm}\\
\ds X_3^\chi =U_2^{-1} x_3 U_2 =x_3 +e_3 \frac{S_{5c} p_c}{E^2}, \qquad
S_{5c} =-\frac 12 i \gamma_c.
\ea
\ee

The position operator on a set of solution $\{\Psi\}$ of eq. (1.1) looks as follows
\be
\ba{l}
\ds X_a^\Psi =U_1^{-1} X_a^\chi U_1 =x_a +e_3 \frac{\gamma_3 S_{5a}}{E}
-e_3 \frac{\gamma_3 S_{5c} p_c p_a}{E^2 (E+|p_3|)} +\frac{S_{ac} p_c}{E(E+|p_3|)},
\vspace{3mm}\\
\ds X_3^\Psi =U_1^{-1} X_3^\chi U_1 =x_3 - \frac{\gamma_3 S_{5c} p_c}{E^2}.
\ea
\ee

(iii) If one performs a transformation on eq. (1.1)
\be
\widetilde U_1 =\frac{1}{\sqrt{2}}(1+\gamma_3)
\ee
and then a transformation
\be
\widetilde U_2 =\frac{E+p_3 +\gamma_a p_a}{\{2E(E+|p_3|)\}^{1/2}},
\ee
it will transform into the equation
\be
i\frac{\p \widetilde \Phi (t,\pbf{x})}{\p t} =\gamma_0 \widetilde \Phi (t,\pbf{x}), \qquad
\widetilde \Phi=\widetilde U_2 \widetilde U_1 \Psi.
\ee

The generators of the group $P(1,3)$ on $\{\widetilde \Phi\}$ coincide with (2.11) where the
sub\-sti\-tu\-tion was made $e_3\to 1$, $|p_3|\to p_3$.

\medskip

\renewcommand{\theequation}{3.\arabic{equation}}
\setcounter{equation}{0}

\centerline{\bf 3. $\pbf{P}$-, $\pbf{T}$- and $\pbf{C}$-properties of two-component equation}

Here we shall study the properties of one of the two-component
eqs. (2.7)\footnote{In what follows, under $\chi$ we shall understand the two-component
spinor $\chi_+$.}
\be
i\frac{\p \chi(t,\pbf{x})}{\p t} =(i\sigma_3 \sigma_a p_a +\sigma_3 |p_3|) \chi(t,\pbf{x}),
\ee
under the discrete transformations.

We shall denote through $P^{(k)}$ ($k=1,2,3$) the space inversion operator of one axis which
is determined as
\be
P^{(1)}\chi(t,x_1,x_2,x_3) =r^{(1)}\chi(t,-x_1, x_2,x_3).
\ee
Analogously $P^{(2)}$ and $P^{(3)}$ are determined.

As is well known, two non-equivalent definitions of the time-reflection operator exist. According
to Wigner the time-inversion operator is
\be
T^{(1)} \chi(t,\pbf{x}) =\tau^{(1)} \chi^*(-t,\pbf{x}).
\ee
According to Pauli it is:
\be
T^{(2)} \chi(t,\pbf{x}) =\tau^{(2)} \chi(-t,\pbf{x}).
\ee

The operator of the charge conjugation can be defined as the product of the operators
$T^{(1)}$, $T^{(2)}$ or as
\be
C \chi(t,\pbf{x}) =\tau^{(3)} \chi^*(t,\pbf{x}),
\ee
where $r^{(k)}$, $\tau^{(k)}$ are the $2\times 2$ matrices.

The operators $P$, $T$, $C$ with the group $P(1,3)$ generators satisfy the usual com\-mu\-ta\-tion
relations.

The generators of the group $P(1,3)$ on the solutions $\{\chi\}$ of eq.~(3.1) have the form
of eq.~(2.5) where
\be
\ba{l}
\ds {\mathcal H}^\chi \to i\sigma_3 \sigma_a p_a +\sigma_3 |p_3| =-\sigma_2 p_1 +
\sigma_2 p_2 +\sigma_3 |p_3|,
\vspace{2mm}\\
\ds S_{ab}\to \frac 14 i (\sigma_b \sigma_a -\sigma_a\sigma_b), \qquad
S_{a3} \gamma_3 \to -\frac 12 \sigma_a,
\ea
\ee
and the matrix $\gamma_0$ is substituted for the matrix $\sigma_3$.

Using the definitions (3.2)--(3.5) it is not difficult to verify that eq.~(3.1) is
$P^{(3)}$-, $C$-invariant but $P^{(1)}$-, $P^{(2)}$-, $T^{(1)}$-, $T^{(2)}$-non-invariant.

Thus, eq. (3.1) is $P^{(3)}C$-, $P^{(1)}P^{(2)}P^{(3)}C$- and
$P^{(a)}C T^{(a)}$-invariant but $P^{(3)}C T^{(a)}$- and $P^{(a)}C$-non-invariant.

We note the following:

(i) The result obtained is a consequence of the fact that the projection operators $Q_\pm$,
with the operators of the discrete transformations, satisfy the following relations
\be
\ba{l}
P^{(a)} Q_\pm =Q_\mp P^{(a)}, \qquad
T^{(a)} Q_\pm =Q_\mp T^{(a)},
\vspace{1mm}\\
P^{(3)} Q_\pm =Q_\pm P^{(3)}, \qquad
C Q_\pm =Q_\pm C.
\ea
\ee

(ii) The two-component equations for the functions $\chi_+$ and $\chi_-$
are equivalent to the four-component one~(2.3) with the subsidiary relativistic-invariant
conditions
\be
Q_- \chi =\left(\frac 12 -i S_{43}\right) \chi =\frac 12 (1-\gamma_3 \gamma_4) \chi=0,
\ee
\be
Q_+ \chi =\left(\frac 12 +i S_{43}\right) \chi =\frac 12 (1+\gamma_3 \gamma_4) \chi=0,
\ee
respectively. For eq. (1.1) these conditions look like
\renewcommand{\theequation}{3.\arabic{equation}${}'$}
\setcounter{equation}{7}
\be
\left(\frac 12 + i e_3 S_{45}\right) \Psi =\frac 12 (1-e_3 \gamma_4)\Psi=0,
\ee
\renewcommand{\theequation}{3.\arabic{equation}${}'$}
\setcounter{equation}{8}
\be
\left(\frac 12 - i e_3 S_{45}\right) \Psi =\frac 12 (1+e_3 \gamma_4)\Psi=0.
\ee

Eq. (1.1) with the subsidiary conditions ($3.8'$) and ($3.9'$) can be joined and can be written
in the form of two $P^{(a)}$- and $T^{(b)}$-non-invariant but $P^{(3)}$- and $C$-invariant equations
\[
\left\{ \gamma_\mu p^\mu +\varkappa (1+e_3 \gamma_4)\right\}\Psi_1(t,\pbf{x})=0,
\qquad
\left\{ \gamma_\mu p^\mu +\varkappa (1-e_3 \gamma_4)\right\}\Psi_2(t,\pbf{x})=0,
\]
where $\varkappa$ is some constant value. The four-component equations for the neutrino, which
are the union of eq.~(1.1) and the usual subsidiary condition, were recently considered in ref.~[6].
These equations, as well as the Weyl equations~(1.2), are $P$- and $C$-non-invariant but
$T^{(1)}$-invariant.

The unitary operator of type $U_2$ for the two-component eq. (3.1) has the form
\renewcommand{\theequation}{3.\arabic{equation}}
\setcounter{equation}{9}
\be
V_1 =\exp \left\{ i\frac{S_a p_a}{\sqrt{p_a^2}} \, \mbox{arctg}\, \frac{\sqrt{p_a^2}}{|p_3|}\right\},
\qquad S_k =\frac 12 \varepsilon_{kln} S_{ln},
\ee
or
\be
V_1 =\frac{E+|p_3|+ i\sigma_a p_a}{\{2E(E+|p_3|)\}^{1/2}}.
\ee

The position operator on the set of solutions $\{\chi\}$ of eqs. (3.1) looks as follows
\be
\ba{l}
\ds X_a^{\chi_+} =V_1^{-1} x_a V_1 =x_a -\frac{\sigma_a}{2E} +
\frac{\sigma_c p_c p_a}{2E^2 (E+|p_3|)}-i\frac{(\sigma_a \sigma_c-
\sigma_c \sigma_a)p_c}{4E(E+|p_3|)},
\vspace{2mm}\\
\ds X_3^{\chi_+} =V_1^{-1} x_3 V_1 =x_3 +e_3 \frac{\sigma_b p_b}{2E^2}.
\ea
\ee

To complete our treatment, we find the position operator for the neutrino which is described
by the Weyl equation~(1.2), for example for the function $\Psi_+$. This equation under a
transformation
\be
V=\frac{E+|p_3|+i\sigma_k \xi_k}{2\sqrt{\xi_k p_k}},
\ee
where the vector $\xi$ has the following components
\[
\xi_k \equiv \left\{ p_1 -p_2 e_3, p_2 +e_3 p_1, e_3(E+|p_3|)\right\},
\]
takes a canonical form
\be
i\frac{\p \Phi_+(t,\pbf{x})}{\p t} =\sigma'_3 E \Phi_+(t,\pbf{x}), \qquad
\sigma'_3 =\sigma_3 e_3, \qquad \Phi_+(t,\pbf{x})= V \Psi_+(t,\pbf{x}).
\ee

The position operator for a neutrino which is described by the Weyl equation~(1.2) (for
$\Psi_+$) looks like
\[
\ba{l}
\ds X_a^W =V^{-1} x_a V =x_a +ie_3 \frac{\sigma_3 \sigma_a}{2E}-i
\frac{e_3 \sigma_3 \sigma_c p_c p_a}{2E^2 (E+|p_3|)} -i
\frac{(\sigma_a\sigma_c -\sigma_c \sigma_a)p_c}{4E(E+|p_3|)},
\vspace{2mm}\\
\ds X_3^W =V^{-1} x_3 V =x_3 -i\frac{\sigma_3 \sigma_b p_b}{2E^2}.
\ea
\]

The other definitions of the operators $X_k$ and $V$ for the neutrino are given in ref.~[5].

(iii) From Dirac eq. (1.1) one can, generally speaking, obtain three types of non-equivalent
two-component equations. On the set of solutions of eq.~(1.1) a direct sum of four
irreducible representations $D^\varepsilon (s)$ of the group $P(1,3)$
\be
D^{\varepsilon=1}\left( s=\frac 12\right) \oplus
D^{\varepsilon=-1}\left( s=-\frac 12\right) \oplus
D^{\varepsilon=1}\left( s=-\frac 12\right) \oplus
D^{\varepsilon=-1}\left( s=\frac 12\right)
\ee
is realized, where $\varepsilon$ is an energy $\mbox{sign}$, $s$ is a helicity. Hence it
follows that there exist three types of two-component equations on the set of which the
following re\-pre\-sen\-ta\-tion of the group $P(1,3)$
\[
D^{\varepsilon=1}\left( s=\frac 12\right) \oplus
D^{\varepsilon=-1}\left( s=-\frac 12\right),
\]
or
\be
D^{\varepsilon=1}\! \left( s=-\frac 12\right) \oplus
D^{\varepsilon=-1}\! \left( s=\frac 12\right), \quad
D^{\varepsilon=1}\! \left( s=\frac 12\right) \oplus
D^{\varepsilon=-1}\! \left( s=\frac 12\right),
\ee
or
\be
D^{\varepsilon=1}\!\!  \left( s=-\frac 12\right) \oplus
D^{\varepsilon=-1}\!\! \left( s=-\frac 12\right), \quad
D^{\varepsilon=1}\!\! \left( s=\frac 12\right) \oplus
D^{\varepsilon=-1}\!\! \left( s=-\frac 12\right),
\ee
or
\be
D^{\varepsilon=1}\left( s=\frac 12\right) \oplus
D^{\varepsilon=-1}\left( s=-\frac 12\right)
\ee
are realized. If on the solutions of two-component equation there realizes the
re\-pre\-sen\-ta\-tion~(3.16) then this equation will be $T^{(1)}$-invariant but
$C$-, $P$-, $T^{(2)}$-non-invariant, if the re\-pre\-sen\-ta\-tion~(3.17) does then it will be
$T^{(1)}$-, $T^{(2)}$-, $C$-invariant but $P$-non-invariant, and if the representation~(3.18)
it will be $T^{(1)}$-, $P$-invariant but $C$-, $T^{(2)}$-non-invariant. This problem will be
considered in more detail in another paper.

\medskip
\renewcommand{\theequation}{4.\arabic{equation}}
\setcounter{equation}{0}

\centerline{\bf 4. Equation for a flat neutrino}

The motion group in the Minkovski three-space is the $P(1,2)$ group of rotations and
translations conserving the form
\[
x^2 =x_0^2 -x_1^2 -x_2^2.
\]
In this case the simplest spinor equation is
\be
i\frac{\p \chi_\pm (t, x_1,x_2)}{\p t} =(i\sigma_3 \sigma_a p_a \pm \sigma_3 m)\chi_\pm (t,x_1,x_2),
\ee
$\chi_\pm$ is the two-component spinor and $m$ is the eigenvalue of the operator
$\sqrt{P_\mu^2}$.

Eq. (4.1) for $\chi_+$ (or $\chi_-$) like eq. (3.1) is invariant under the $P^{(1)} P^{(2)}$-
and $C$-operations but non-invariant under the $P^{(a)}$ and $T^{(b)}$-operations.

Thus, eq. (4.1) for the wave function $\chi_+$ (or $\chi_-$) is $P^{(1)}P^{(2)}C$-,
$T^{(a)}P^{(b)}$- and $P^{(a)} CT^{(b)}$-invariant but $P^{(a)}C$- and $C T^{(a)}$-non-invariant.

It should be noted that the equation being the ``direct sum'' of the equation for
$\chi_+(t,x_1,x_2)$ and $\chi_-(t,x_1,x_2)$ is invariant under the $P$-, $T$- and
$C$-transformations~[7].

Finally, we quote one more example of the $P$- and $C$-non-invariant equation which is
invariant with respect to the inhomogeneous De Sitter group. Such is the Dirac equation:
\be
i\frac{\p \Psi(t,\pbf{x}, x_4)}{\p t} =(\gamma_0 \gamma_k p_k +\gamma_0 \varkappa)
\Psi(t,\pbf{x}, x_4), \qquad k=1,2,3,4.
\ee
This equation as is shown in refs. [2, 7] is $T^{(1)}$-, $T^{(2)}C$-invariant but
$P^{(k)}$-, $T^{(2)}$- and $C$-non-invariant.

All the results obtained in this paper can be generalized for the arbitrary spin $s$ case, if
one uses for this the purpose the equation (ref.~[2]):
\be
i\frac{\p \Psi(t,\pbf{x})}{\p t} =\lambda S_{0l} p_l \Psi(t,\pbf{x}), \qquad l=1,2,3,
\ee
where $\lambda$ is some fixed parameter (for the Dirac equation $\lambda=-2i$),
and $S_{\mu\nu}$, $S_{\mu 4}$, $S_{45}$ are the
matrices (not $4\times 4$ ones) realizing the algebra $O(1,5)$ representation.

(i) If we transform the usual Dirac equation describing the motion of the non-zero mass particle
$m$ with a spin $\frac 12$ as
\be
V_2 =\frac{\gamma_3 p_3 +q_3 +m}{\{2q_3 (q_3+m)\}^{1/2}},
\qquad q_3 \equiv \sqrt{p_3^2 +m^2},
\ee
it has the form
\be
i\frac{\p \Psi'(t,\pbf{x})}{\p t} =H' \Psi'(t,\pbf{x}),
\ee
\be
H' =\gamma_0 \gamma_a p_a +\gamma_0 q_3, \qquad \Psi'=V_2 \Psi, \qquad a=1,2.
\ee
Choosing the representation (2.6) for the Dirac matrices eq. (4.5) is decomposed into the set
of two independent equations
\be
i\frac{\p \Psi'_+(t,\pbf{x})}{\p t} =(-\sigma_2 p_1 +\sigma_1 p_2 +\sigma_3 q_3 )\Psi'_+(t,\pbf{x}),
\ee
\be
i\frac{\p \Psi'_-(t,\pbf{x})}{\p t} =(-\sigma_2 p_1 +\sigma_1 p_2 -\sigma_3 q_3 )\Psi'_-(t,\pbf{x}),
\ee
where $\Psi'_+$ and $\Psi'_-$ are two-component wave functions.

Eq. (4.7) or (4.8) describes a free motion of spinless particle and antiparticle with the mass $m$.
Thus besides of the Klein--Gordon equation there exist the other equations of the type~(4.7)
and~(4.8) which are also relativistically invariant and describe the spinless particle motion
with non-zero mass. The two-component eq.~(4.7) is equivalent to the four-component
Dirac equation
\be
i\frac{\p \Psi(t,\pbf{x})}{\p t} = (\gamma_0 \gamma_k p_k +\gamma_0 m) \Psi(t,\pbf{x}),
\qquad k=1,2,3
\ee
with such subsidiary condition
\be
\left( 1-\frac{\gamma_3 \gamma_4 m+ \gamma_4 p_3}{q_3} \right) \Psi(t,\pbf{x})=0.
\ee

The author expresses his gratitude to Professor O.S.~Parasjuk for the benefits of
many discussions.

\medskip

\begin{enumerate}
\footnotesize

\item Lee T.D., Yang C.N., {\it Phys. Rev.}, 1957, {\bf 105}, 1671; \\
Landau L., {\it Nucl. Phys.}, 1957, {\bf  3}, 127; \\
Salam A., {\it Nuovo Cimento}, 1957, {\bf  5},  1207.

\item Fushchych W.I., Kiev, preprint ITF-70-40, 1970. {\tt quant-ph/0206079}

\item Shirokov Yu.M., {\it JETP (Sov. Phys.)}, 1958, {\bf  6}, 664.

\item Lomont J.S., Moses H.E., {\it J. Math. Phys.}, 1962, {\bf 3}, 405.

\item Fronsdal C., {\it Phys. Rev.}, 1959, {\bf  113}, 1367;\\
Voisin J., {\it Nuovo Cimento}, 1964, {\bf 34},  1257.

\item Tokuoka Z., {\it Prog. Theor. Phys.}, 1967, {\bf  37}, 603;\\
Sen Gupta N.D., {\it Nucl. Phys. B}, 1968, {\bf 4}, 147; \\
Santhanam T.S., Chandrasekaran P.S., {\it Prog. Theor. Phys.}, 1969, {\bf 41}, 264.

\item Fushchych W.I., Kiev, preprint ITF-69-17, 1969.\ \ {\tt quant-ph/0206045}

\end{enumerate}
\end{document}